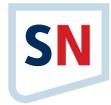

# DFENet: A Novel Dimension Fusion Edge Guided Network for Brain MRI Segmentation

Hritam Basak[1] · Rukhshanda Hussain[1] · Ajay Rana[2]

**Abstract**

The rapid increment of morbidity of brain stroke in the last few years have been a driving force towards fast and accurate segmentation of stroke lesions from brain MRI images. With the recent development of deep-learning, computer-aided and segmentation methods of ischemic stroke lesions has been useful for clinicians in early diagnosis and treatment planning. However, most of these methods suffer from inaccurate and unreliable segmentation results because of their inability to capture sufficient contextual features from the MRI volumes. To meet these requirements, 3D convolutional neural networks have been proposed, which, however, suffer from huge computational requirements. To mitigate these problems, we propose a novel Dimension Fusion Edge-guided network (DFENet) that can meet both of these requirements by fusing the features of 2D and 3D CNNs. Unlike other methods, our proposed network uses a parallel partial decoder (PPD) module for aggregating and upsampling selected features, rich in important contextual information. Additionally, we use an edge-guidance and enhanced mixing loss for constantly supervising and improvising the learning process of the network. The proposed method is evaluated on publicly available Anatomical Tracings of Lesions After Stroke (ATLAS) dataset, resulting in mean DSC, IoU, Precision and Recall values of 0.5457, 0.4015, 0.6371, and 0.4969 respectively. The results, when compared to other state-of-the-art methods, outperforms them by a significant margin. Therefore, the proposed model is robust, accurate, superior to the existing methods, and can be relied upon for biomedical applications.

✉ Hritam Basak
  hritambasak48@gmail.com

  Rukhshanda Hussain
  rukhshanda189@gmail.com

  Ajay Rana
  ajay.rana6288@gmail.com

[1]  Department of Electrical Engineering, Jadavpur University, 188, Raja S.C. Mullick Road, Jadavpur, Kolkata, West Bengal 700032, India

[2]  Department of Computer Science Engineering, SRM Institute of Science and Technology, SRM Nagar, Kattankulathur, Chennai, Tamil Nadu 603203, India

## Introduction

Cerebrovascular accident is among the most common diseases, prevailing amongst the community in age groups of 40–60 contributing to a huge percentage of death worldwide every year [11]. Studies have shown that they can even cause disabilities in adults for 2–5 years in about 37–71% of reported cases globally [31]. Rehabilitation may be constructive for an eventual recovery in acute conditions with its effectiveness depending on the neurological developments and damages caused by a stroke in the patients. However, significant improvements in neuro-imaging including the brain image analysis and the T1-weighted magnetic resonance imaging (MRI) images have proven to be contributory for researchers in the diagnosis of patients, improvements in treatment procedures or the likelihood of gaining back some functionality like the motor speech [28]. Brain MRI segmentation is an important task because it influences the entire diagnosis with the processing steps depending on accurate segmentation of various anatomical regions. MRI combined with other diagnostic procedures thus helps in the detection of minor strokes and the presence of Ischemia that may result in Ischemic strokes. In recent days, convolutional neural networks abbreviated as CNN and deep neural networks (DNN) models have proven to be remarkably convenient for classification and segmentation purposes such as [5, 14, 18]. These methods differ from classical image processing methods in several aspects including the automated feature extraction framework [7]. Deep-learning based methods, as compared to the traditional ones, are efficient, robust and effective for clinical applications [6]. The 2-D CNN is used to convert the volumetric data of MRI images into two-dimensional slices, predicting the result. With each iteration, the network improves its result by minimizing these losses. However, as considerable spatio-temporal information is obscured in this approach as shown in the recent studies, the researchers have been shifting towards 3-D CNNs.

3-D CNNs are trained to utilize this crucial information in volumetric data and to segregate and outline any abnormalities in the domain of medical imaging. Nonetheless, the memory and computational requirements used for 3-D CNNs are difficult to meet. Hence, they have been mostly avoided even though they might be useful to extract important volumetric information. The U-net architectures have been quite instrumental in present days for the stated purpose [3, 27, 34], however, they are unable to extract information through 2D operations from 3D MRI data. For solving the problem in segmentation of the areas of stroke-lesion precisely, in this paper we propose a novel dimensionally fused U-net framework which, despite having a 2-D framework, can also associate with relevant spatial 3-D information besides the 2-D information with considerably lower resources as far as memory requirements and dataset are concerned. Besides, instead of generating a segmentation map through the traditional upsampling layers, we propose a parallel partial decoder (PPD) module, that aggregates the features from different layers of the CNN for a better saliency map. It takes in the contextual information to generate a global segmentation map acting as the prediction of the proposed network.

MRI images often contain regions containing a certain level of similarity between two adjacent classes (foreground and background). Hence, accurate prediction of class boundary is somehow difficult for regions having higher-level semantic information from both of the classes. To mitigate this problem, an edge attention module is associated for better representation of the respective edge information of the stroke lesions. Also, we propose an enhanced mixing loss function that combines the standard Binary Cross Entropy (BCE) loss function along with the weighted Intersection over Union (IoU) loss function that helps in enhancing the gradient propagation and achieving faster convergence. As a result of this enhanced deep supervision, the proposed model learns sufficient and important gradient information of pixel intensity and lesion boundary, leading to an accurate and improved segmentation map.

## Overview and Contribution

This paper proposes a novel segmentation framework that utilizes both 3D and 2D CNN information for accurate segmentation of stroke lesion. The contribution of the paper is described as follows:

1. We propose a novel dimension fusion block to integrate the features from 2D and 3D CNNs for better spatio-temporal feature representation. The proposed model is lightweight as compared to 3D CNNs, providing superior performance than 2D CNNs.
2. Instead of simple upsampling layers, we propose a parallel partial decoder (PPD) module to associate features from deep layers of CNN, containing high-level semantic information for a better saliency map.
3. The shallow features obtained from the first convolution layer of the CNN is used for edge guidance for accurate mapping of image regions near lesion boundary.
4. We propose an enhanced mixing loss function integrating the traditional weighted IoU and BCE loss function for better supervision and faster convergence.

## Literature Survey

Recent researches in the literature mostly address the problem of lesion segmentation from brain MRI as a semantic segmentation by producing dense pixel-wise predictions for every slice of image. Quite noteworthy outcomes have been attained by researchers using handcrafted features for brain MRI segmentation purposes in the last few years [33]. A multivariate CTP based segmentation technique of MRI images with a chance of infarct voxelwise was presented by Kemling et al. [19] whereas Nabizadeh et al. [26] proposed a histogram improvement algorithmic law using DWI for ischemic lesion segmentation. A multimodal MRI image localization method was then projected by Mitra et al. [25] for the feature mining followed by a Random Forest method for the lesion segmentation which decreased the false positive rate considerably.

Recent advancements in deep learning algorithms have inspired researchers to effectively utilize CNN based approach for supervised lesion segmentation. UNet [32] is often considered as the state-of-the-art for segmentation of biomedical images, however several modifications have been proposed in recent years to improve the overall learning and feature representation [2, 4, 20, 21]. Lyksborg et al. [24] used a 3-path ensemble of CNN networks, each for the canonical axial, sagittal and coronal views. Chen et al. [9] proposed a novel ensemble approach of multi-scale convolutional label evaluation net (MUSCLE Net) and DeconvNet that outperformed existing methods on a private MRI dataset. Badrinarayanan et al. [1] proposed SegNet architecture, consisting of an encoder–decoder architecture followed by a pixel wise classification layer. The proposed model, when compared with widely adopted segmentation frameworks, was proven to be exceptionally better. Multi-path U-Net was proposed by Dolz et al. [10] initially to address the variability of ischemic strokes' location and shape, but later was adopted in several other medical image segmentation tasks. However, all these methods fail to extract the important 3D context information in the 3D volume data, thus the prediction might lose continuity due to the limitations of 2D slices.

To address the shortcomings of these methods, 3D CNNs have been proven to be of great potential recently. Fully connected 3D DenseNet was proposed by Zhang et al. [36] with comparatively deep architecture and tight connections for improved performance. A two-path 3D CNN structure consisting of fully connected 3D conditional random field was proposed by Kamnistas et al. [18] which was implemented on ISLES2015 dataset with competitive results. A 3D Seg-Net architecture was proposed by Hu et al. [15] consisting of 3D residual framework with 3D voxel-wise segmentation pipeline. The proposed method produced promising results on ISLES2017 dataset. Later Feng et al. [12] proposed a method that extracts both spatial and temporal information using 3D convolution operations, and is able to capture important dynamic semantic features from adjacent frames. However, all these models require huge computational cost and very long time to train. Therefore it is very difficult to fine-tune the training hyperparameters and therefore, this models are prone to overfitting for small datasets. As a remedy, models with both 2D and 3D methods, connected in cascaded manner, were proposed. A hybrid densely connected U-Net (HDenseU-Net) was proposed by Li et al. [21] where a 2D densely connected U-Net was used for initial segmentation, followed by a 3D CNN for correction of spatial continuity. However, these methods often suffer from loss of contextual information of fine-grained boundary regions through a series of downsampling and pooling operations in the case of deep CNNs. Secondly, training traditional CNN models requires lots of labelled data which is often not present in brain MRI dataset. These challenges lead to shallow networks with intelligent orientations of layers, that requires fewer parameters and small contextual information for superior performance.

## Proposed Method

In this section, we explain the workflow of our proposed method, the network architecture, different modules and blocks used, the proposed loss function and the hyperparameters involved.

### Feature Extraction and Aggregation

The primary architecture of the proposed network is built upon the U-Net as the base model with a few additional modifications. The overall framework has been developed to extract both high-level semantics and the low-level surface information from the image data in the architecture presented in Fig. 1. The proposed network consists of two parallel branches used for extraction of different dimensional spatial information from the volumetric MRI dataset through a series of 2D and 3D convolutions, and later merged using a dimension transfer block, as shown in Fig. 2. This fusion scheme enables the network learn refined edge information and facilitates the model identify small stroke regions. As the network deepens, the total number of trainable parameters increases extremely due to the 3D convolution operations. Hence, we have used the fusion scheme only in the early stages of the network.

The dimension transfer block performs three major operations: (1) 3D dimensionality reduction, (2) squeeze and excitation operation, and (3) feature fusion. The idea was adopted from the recently developed Squeeze and Excitation (SE) network by Hu et al. [16]. The parameter $r$ inside the SE block is defined as the reduction ratio that regulates the computational cost and capacity of the block. This special architectural block specifically activates the channel-wise dynamic re-calibration of 3D and 2D feature branches and enhance the fusion effect of two different dimensional features. The excitation part of the SE block is the weighting of every feature map present in the side network but at a relatively lower computational cost. Thus, even smaller regions of stroke lesions, considered of utmost importance in various applications of medicine, can be easily detected following the given approach.

Let $F_{2d}$ and $F_{3d}$ are used to denote the 2D and 3D feature maps respectively, which are the inputs for the dimension transfer block, whilst the depth, width, height, the number of channels and the batch size of feature map are represented by $D$, $W$, $H$, $C$, $N$ respectively. Firstly there is a conversion in the dimension of the $F_{3d}$ from $N \times H \times W \times D \times C$ to $N \times H \times W \times D \times 1$ to produce intermediate feature $F *_{3d}$ which is done with the help of the convolution operation utilizing a 3D $1 \times 1 \times 1$ convolutional block. Following this, there is a reduction in the dimensionality of $F *_{3d}$ from $N \times H \times W \times D \times 1$ to $N \times H \times W \times D$ using squeeze block. However, to maintain the consistency in dimension with the two-dimensional feature maps, the $F *_{3d}$ is converted from having a dimension of $N \times H \times W \times D$ to $N \times H \times W \times C$ by applying a 2D $3 \times 3$ convolutional block setting the filter size to $C$. $f_d$ is used to denote the dimensionality reduction operation function where the conversion of $F_{3d}$ to $F'_{3d}$ is done on application of the stated function (where $F'_{3d}$ is having a dimension of $N \times H \times W \times C$):

$$F'_{3d} = f_d(F_{3d}), \tag{1}$$

where $f_d$ represents the dimensionality reduction operation.

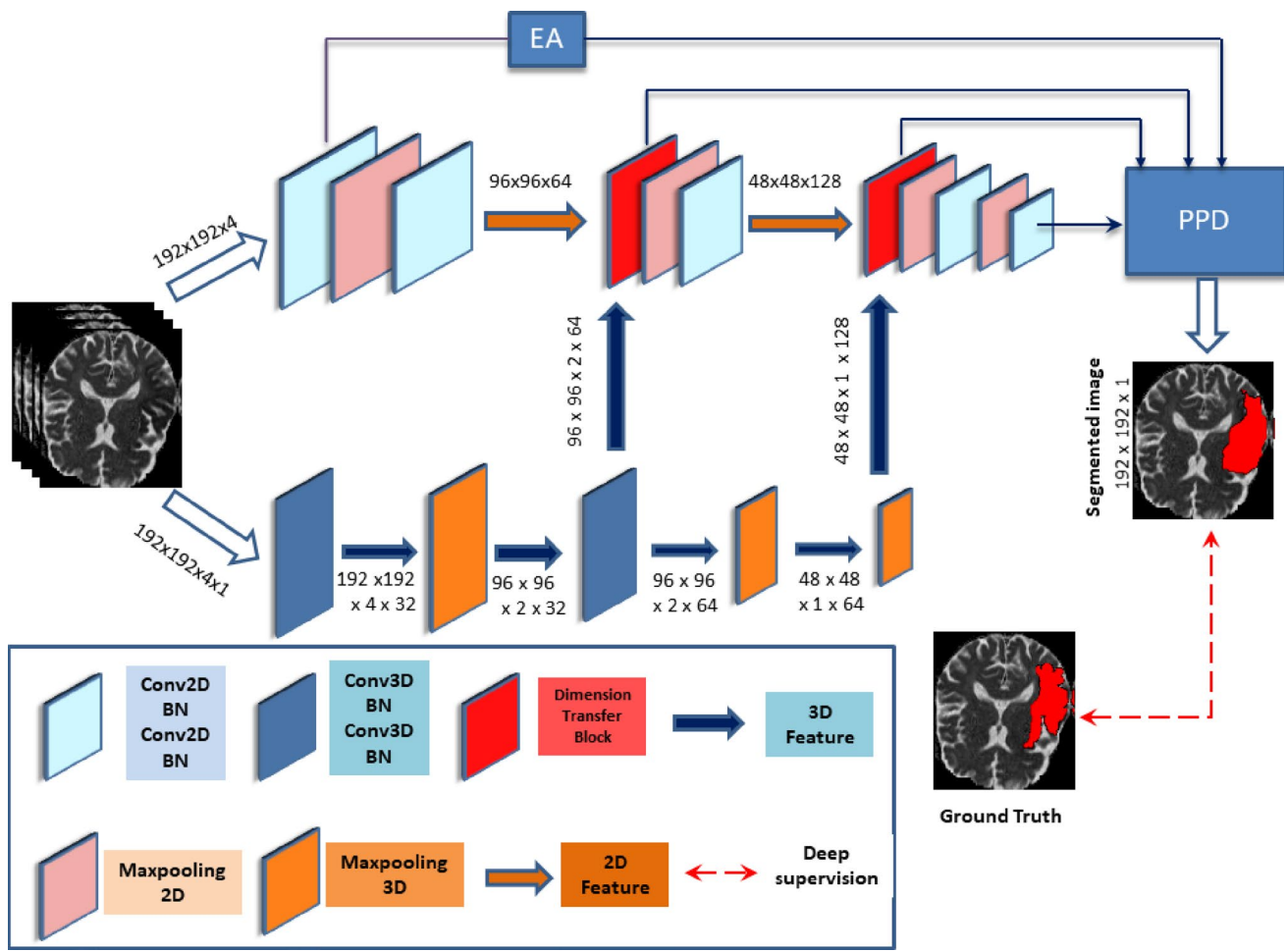

**Fig. 1** The overall workflow of the proposed DFENet

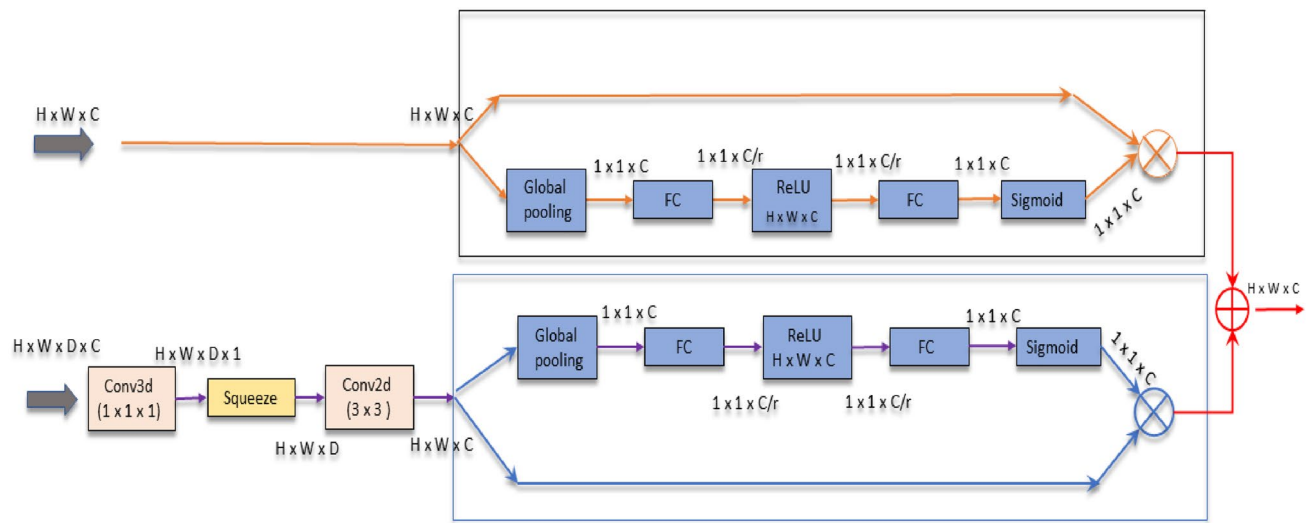

**Fig. 2** Architecture of the Dimension transfer block, with two branches as input from two parallel 2D and 3D networks. First the feature channel of 3D network is compressed to 1 followed by a squeeze block and 2D $3 \times 3$ convolution operation, resulting in the feature dimension consistent with the 2D feature branch. Finally the features are passed through SE block where $r$ represents the reduction ratio. Finally the two branches are merged together to form the fused output

Next, the SE block is used to aggregate the feature channels of two different dimensions where channel outputs are weighted and thus considered for better proficiency in feature expression before fusion. Mathematically, $F$ (where $F$ is having a dimension of $N \times H \times W \times C$) is denoting a fused feature map that is obtained in the form of an output of the SE block. Further parameters existing in the entire network and its architecture are described in Fig. 2. In this step, the features from two different dimensions are fused, where $f_s$ is representing the SE block that squeezes as well as excites.

$$F = f_s(F'_{3d}) + f_s(F_{2d}) \tag{2}$$

## Edge Attention

Accurate segmentation of regions near lesion boundary from biomedical images is quite challenging due to the high-level semantic information shared between two adjacent classes (foreground and background) near lesion boundaries. Zhao et al. [39] proposed that fine-grained boundary constraints can provide useful supervision over the feature extraction task for image segmentation, leading to accurate localization of ROI. Later, similar claims were made by Et-Net [38] here the authors utilized edge attention representations in the early encoding stage and later transferred them to multi-stage decoding layers for biomedical image segmentation. Hence, being inspired by the original work of DANet [13] we introduce a supervised edge attention module to the segmentation framework to effectively learn the edge information from the instances. Here position attention module, as suggested by Ref. [13], enhances the representation capability of the local features by aggregating a wide range of contextual information into them. In addition to the original DANet, we add additional convolution layers after the position attention module to obtain edge attention features of the same depth as the feature.

Let $F_A \in \mathbb{R}^{h \times w \times c}$ be a local feature, where $h$, $w$, and $c$ represent the height, width and channel respectively, be fed to three convolution layers to produce three new feature maps $F_B$, $F_C$ and $F_D$, where $\{F_B, F_C, F_D\} \in R^{h \times w \times c}$. $F_B$, $F_C$ and $F_D$ are then reshaped to $n \times c$, where $n = h \times w$ represents the number of pixels. Then $F_B$ is multiplied with transpose of $F_C$, followed by a *softmax* to obtain edge attention map $\mathcal{M}_E \in \mathbb{R}^{n \times n}$.

$$\mathcal{M}_E(j, i) = \frac{\exp\left(F_B^i \cdot F_C^j\right)}{\sum_{i=1}^{n} \exp\left(F_B^i \cdot F_C^j\right)} \tag{3}$$

where $\mathcal{M}_E(j, i)$ represents the impact of the $i$th position pixel's on that of the $j$th position's. In parallel, $F_D$ is multiplied with transpose of $\mathcal{M}_E$ and the output is reshaped to $h \times w \times c$. Finally, a multiplication factor $\beta$ is multiplied to it and the result is element-wise summed up with $F_A$ to produce the final output of EA module $O_{EA}$, given by the following equation:

$$O_{EA}(j) = \beta \sum_{i=1}^{n} \left[F_A(i) + \{F_D(i) \times \mathcal{M}_E(j, i)\}\right]. \tag{4}$$

The output from the EA module is then fed to the PPD block for better edge supervision of the overall segmentation process. The workflow of the EA block is shown in Fig. 3.

## Parallel Partial Decoder Module

Traditional segmentation models in the U-Net family utilizes symmetrical encoder-decoder architecture, providing similar importance to high-level and low-level features to produce the final segmentation map. However, [35] suggested that low-level features contribute very little towards the final prediction map, leading to unnecessary usage of computational resource due to their high spatial resolution. To mitigate this problem, instead of aggregating features from all level of CNN followed by sequential upsampling, we have used a partial decoder module that utilizes inputs from selected CNN layers only. Being inspired by the Receptive Field Block (RFB) [23], our proposed PPD module captures global contextual information to produce an accurate segmentation map.

The initial convolution layers of the CNN are considered the shallow layers and provide information with very little significance towards the final prediction. Hence, they are discarded from the inputs of the PPD module. Specifically, the features from the two-dimension transfer blocks are considered as the major inputs of the PPD module because of their richness in important spatio-temporal information, both in 2D and 3D space. Additionally, the feature from the final downsampling layer and the output from the EA module is also fed to the PPD module. To accelerate the feature propagation, we add a series of convolution and batch normalization operations as shown in Fig. 4. Short skip connections are added in the PPD module, similar to the original RFB module. After obtaining different discriminating features from different layers, we finally multiply them to reduce the gap between multiple feature levels. Thus, the PPD module produces a global segmentation map through a series of element-wise multiplication and concatenation operations. The overall architecture of the PPD module is shown in Fig. 4.

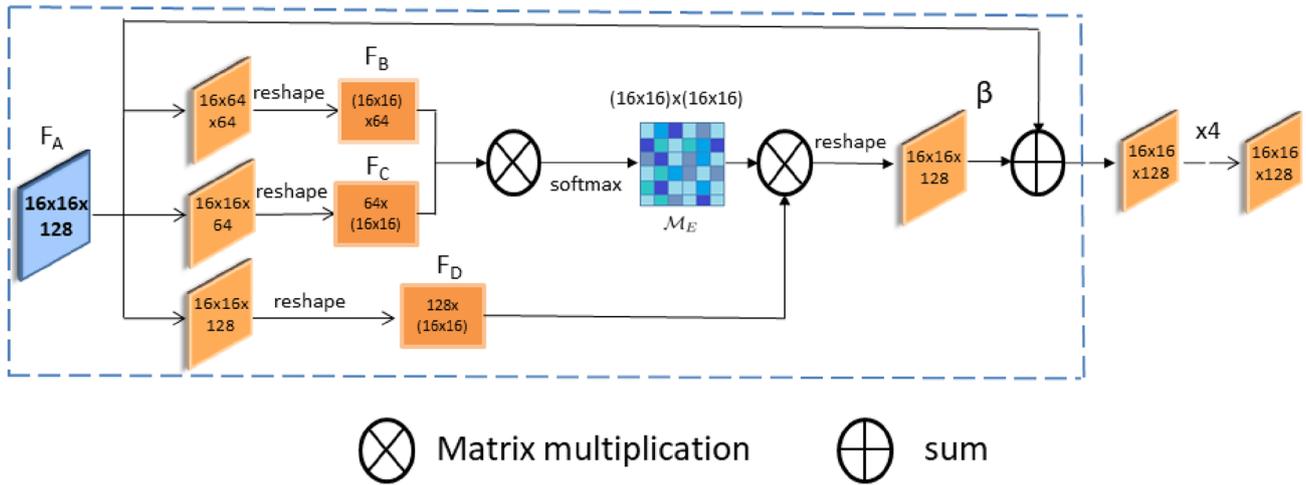

**Fig. 3** The workflow of the proposed EA module

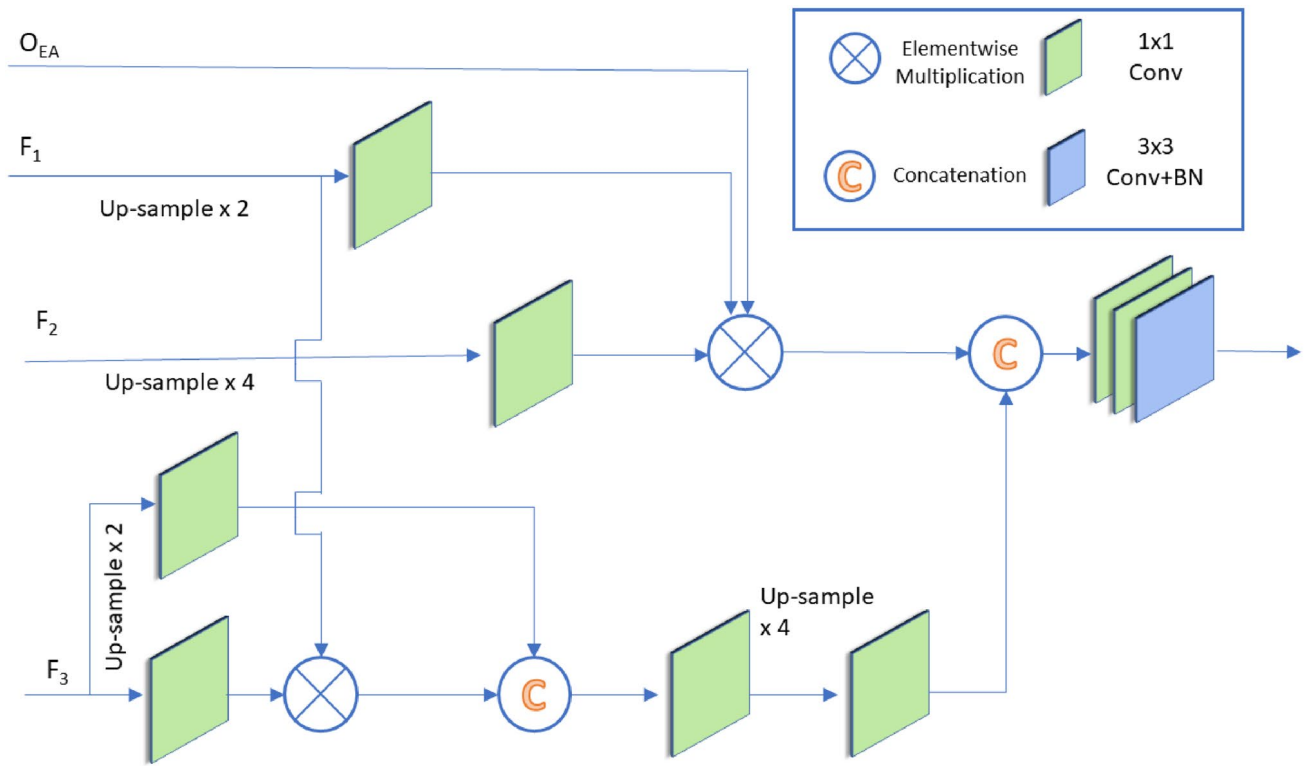

**Fig. 4** The proposed Parallel Partial Decoder (PPD) module that utilizes three different feature inputs from different CNN layers and edge attention output to generate the global segmentation map. Here $O_{EA}$ represents output from EA module, $F_1$ and $F_2$ represent features from dimension transfer block-1 and 2 respectively, $F_3$ is the feature from final downsampling block

## Enhanced Mixing Loss

To supervise the overall learning and segmentation process of the proposed network, we have used an enhanced mixing loss function to mitigate the problems of standard Least Absolute Deviations ($L1$) and Least Square Errors ($L2$) loss functions. This hybrid loss function incorporates the features of weighted Binary CrossEntropy (BCE) and weighted Intersection over Union (IoU) loss functions, leading to fast convergence and efficient global and local supervision.

The result from the EA module $O_{EA}$ is supervised with respect to the actual edge map $G_E$, obtained by calculating the gradient of the ground truth. The standard BCE loss function is used to calculate the dissimilarity between these two as follows:

$$\mathcal{L}_{EA} = -\sum_i G_E \log(O_{EA}) + (1 - G_E) \log(1 - O_{EA}), \quad (5)$$

where $i$ indicates the pixel value of the segmentation map.

The weighted BCE loss function, commonly known as WCE [29], is a modification of standard BCE loss, useful in biomedical application where class imbalance between foreground and background pixel is evident. The formulation of WCE loss in our case is as follows:

$$\mathcal{L}_{WBCE} = -\sum_i \beta \times G \log(O_S) + (1 - G) \log(1 - O_S), \quad (6)$$

where $\beta$ is the correcting factor, used to tune the false positive and false negative predictions, $G$ is the ground truth, $O_S$ being the prediction map. Similarly, we calculate the weighted IoU loss function $\mathcal{L}_{WIoU}$ and define the overall loss function $\mathcal{L}$ as:

$$\mathcal{L} = \mathcal{L}_{WIoU}(O_S, G) + \delta \times \mathcal{L}_{WBCE}(O_S, G) + \mathcal{L}_{EA}(O_{EA}, G_E). \quad (7)$$

# Results and Discussions

In this section, we describe the results obtained, both quantitatively and qualitatively, the experimentations performed, ablation studies and the comparison of our results with those of the state-of-the-art, to evaluate the comparative performance of the proposed method.

## Evaluation Metrics

To evaluate the performance of the proposed model on the supervised stroke lesion segmentation task, we have utilized four standard and widely used evaluation metrics, described as follows:

### Dice Similarity Coefficient (DSC)

It is a spatial overlap metric that is computed as in Eq. 8 for predicted image $S$ and ground truth $G$.

$$\text{DSC}(S, G) = \frac{2 \times |S \cap G|}{|S| + |G|}. \quad (8)$$

### Intersection over Union (IoU)

Also known as Jaccard Index (JI), it measures the accuracy of segmentation by computing the ratio of the intersection of objects and their union when projected on the same plane. Mathematically it is expressed as in Eq. 9, where $S$ is the predicted segmentation mask and $G$ is the original ground truth mask of the image.

$$\text{IoU}(S, G) = \frac{|S \cap G|}{|S \cup G|}. \quad (9)$$

### Precision

It adequately refers to the unadulterated positive detections concerning the actual ground truth. Precision addresses how many of the pixels in the segmentation map matches the verified ground truth observations. Mathematically it is expressed as in Eq. 10, where $TPs$ and $FPs$ are the true positive and the false positive respectively.

$$\text{Precision} = \frac{TP}{TP + FP}. \quad (10)$$

False Negative error arises if a pixel inside a stroke region is misclassified as a normal region. Likewise in case of False Positive, a pixel belonging to non-lesion class is misclassified into lesion class.

### Recall

It effectively expresses, how complete the positive predictions are concerning the actual ground truth. Amongst the total pixels annotated in the verified ground truth, it refers to the number of pixels that were captured as positive predictions. Mathematically it is expressed as in Eq. 11, where $TP$ and $FN$ indicate true positive and false negative respectively.

$$\text{Recall} = \frac{TP}{TP + FN}. \quad (11)$$

## Implementation

The proposed method was implemented in Python, utilizing Nvidia K80 GPU with 12 GB available RAM. A Stochastic Gradient Descent (SGD) optimizer with reduced learning is utilized where the learning rate is reduced by a factor of 0.1 on the performance metrics plateaus on the validation set, whereas the initial learning rate was set to $1e-4$. The proposed method was evaluated on Anatomical Tracings of Lesions After Stroke (ATLAS) [22] dataset, consisting of 229 T1 weighted 3D MRI images for stroke lesion segmentation, collected from 11 cohorts worldwide. Each of the images consists of 189 slices, manually segmented by expert practitioners for stroke lesions. The original images have dimensions of $233 \times 197$, resized to $192 \times 192$. Several image augmentation methods including random flip, random rotate, colour augmentation methods were incorporated to address the possibility of over-fitting. The reduction factor $r$ of the SE block is set to 16, the batch size is set to 8, maximum iteration is set to 200, early stopping is included. The dataset was divide into train-test-validation split in a ratio of 8:1:1. 5-fold cross validation was performed and the average value was reported throughout the experiments. Figure 5 represents the variation of different segmentation metrics with respect to epochs, both for training and validation sets. We have shown the mean value of 5 folds as well as $\pm 1\%$ standard deviation value in the figure.

## Ablation Study

To study the importance of different modules of the proposed DFENet, we have performed ablation studies by

**Table 1** The results obtained from the ablation studies on the ATLAS dataset to evaluate their importance

| Instances | Combination | Result | | | |
|---|---|---|---|---|---|
| | | DSC | IoU | Precision | Recall |
| 1 | 2D UNet (base model) | 0.4606 | 0.3447 | 0.5994 | 0.4449 |
| 2 | 3D UNet (base model) | 0.4871 | 0.3664 | 0.6027 | 0.4506 |
| 3 | 2D+3D UNet (using SE) | 0.5094 | 0.3787 | 0.6061 | 0.4773 |
| 4 | 2D+3D UNet+PPD | 0.5383 | 0.4029 | 0.6297 | 0.4821 |
| 5 | 2D+3D UNet+EA | 0.5281 | 0.4003 | 0.6139 | 0.4833 |
| **Proposed** | **2D+3D UNet+EA+PPD** | **0.5457** | **0.4015** | **0.6371** | **0.4969** |

removing a particular module or part of the architecture, keeping all the other learning parameters and network unchanged. Table 1 shows the experimental analysis of this study, leading to the conclusive decisions of the importance of EA, PPD and SE blocks. These blocks, as shown from the table, when integrated with the base model, has boosted the mean DSC and IoU scores from 0.4606 (UNet) to 0.5481 (proposed) and from 0.3447 (UNet) to 0.4015 (proposed) respectively. It is evident from the table that 2D and 3D UNet can produce not so accurate results on the segmentation dataset, whereas combining the features of the former one with the previous one can boost up the performance (instance 3 in Table 1) by aggregating sufficient contextual and spatio-temporal information, which can not be explored using mere 2D UNet. Replacing the upsampling block of UNet with the proposed PPD can effectively increase the

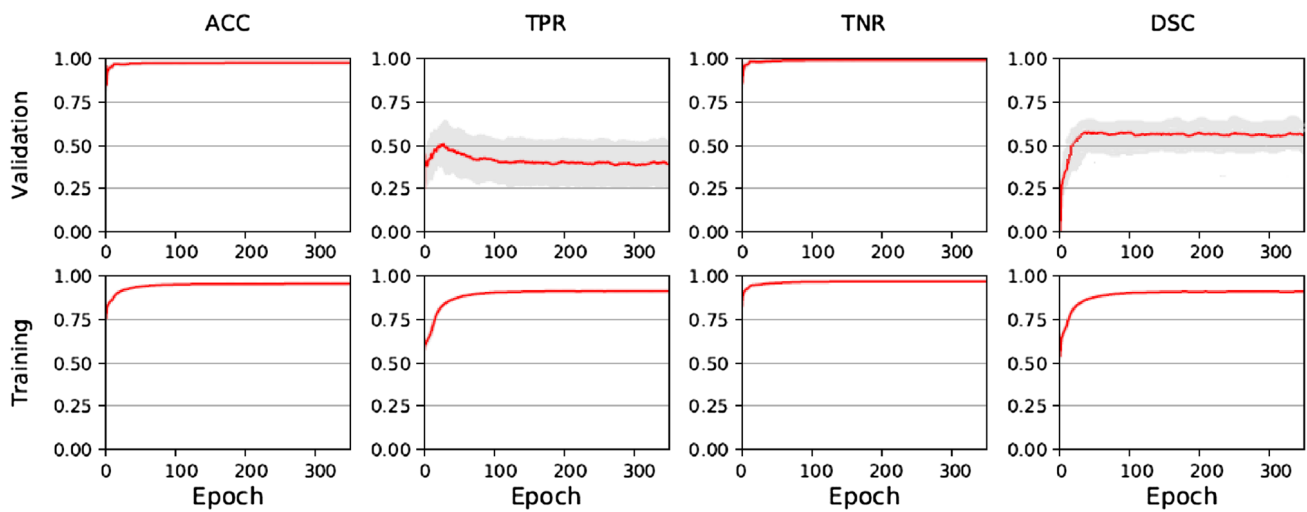

**Fig. 5** Variation of segmentation metrics for validation and train subjects with increasing epochs. The graphs represents the mean values of 5 folds and the gray area illustrates the $\pm 1\%$ standard deviation.

Here '*TPR*' and '*TNR*' represents the true positive and true negative rates respectively, '*DSC*' represents dice similarity score performance by effectively capturing long-range dependencies of the network and discarding the shallow features that contribute less towards the final segmentation output, as shown in instance 4. The shallow level feature, however, if effectively used through an edge attention module, can also make a significant impact towards the final prediction, as shown in instance 5 of Table 1. The last row depicts the overall performance of the proposed DFENet on the ATLAS dataset.

## Comparison with State-of-the-Art

To validate the effectiveness of the proposed DFENet, we have compared our results quantitatively with several other existing methods that have been successfully used for accurate segmentation of stroke lesions on the ATLAS dataset. Table 2 depicts this comparison of our results with UNet [32], SegNet [1], ResUNet [37], PSPNet [40], DeepLab V3 [8], XNet [30], 3D CNN+CRF [18], and Brain SegNet3D [15]. To analyze the visual results of the superiority of the proposed method over most of the existing ones, the visual comparison of some selected segmentation instances are shown in Fig. 6. It is evident from the figure that, SegNet often misses important and minor lesion regions (row 1), leading to poor overall performance as shown in Table 2. Though UNet performs quite well in other biomedical segmentation applications and is considered as a state-of-the-art model, it fails to detect sufficient information for accurate lesion segmentation from the ATLAS dataset as compared to the proposed DFENet, shown both in Tables 1 and 2. XNet and ResUNet on the other hand consistently produced a segmentation map, very close to the proposed DFENet, which is also reflected in Table 2. The 3D segmentation frameworks like 3D CNN+CRF [18] and Brain SegNet3D [15] outperforms the proposed method marginally, as shown in Table 2. But, these methods are computationally expensive as compared to the proposed method. Thus we can conclude that the proposed DFENet outperforms all the 2D segmentation methods in this particular segmentation task and produces comparable results to 3D CNNs with less computational cost.

**Table 2** Comparison of the quantitative results of the proposed DFENet with several other existing methods on the ATLAS dataset

| Dataset | Method | Result | | | |
|---|---|---|---|---|---|
| | | DSC | IoU | Precision | Recall |
| ATLAS | UNet [32] | 0.4606 | 0.3447 | 0.5994 | 0.4449 |
| | SegNet [1] | 0.2767 | 0.1911 | 0.3938 | 0.2532 |
| | ResUNet [37] | 0.4702 | 0.3549 | 0.5941 | 0.4537 |
| | PSPNet [40] | 0.3571 | 0.2541 | 0.4769 | 0.3335 |
| | DeepLab V3 [8] | 0.4609 | 0.3458 | 0.5831 | 0.4491 |
| | XNet [30] | 0.4867 | 0.3723 | 0.6 | 0.4752 |
| | Brain SegNet3D [15] | 0.5632 | 0.4325 | 0.6441 | 0.4852 |
| | 3D CNN+CRF [18] | 0.5667 | 0.4423 | 0.6456 | 0.5410 |
| | **DFENet (Proposed)** | **0.5457** | **0.4015** | **0.6371** | **0.4969** |

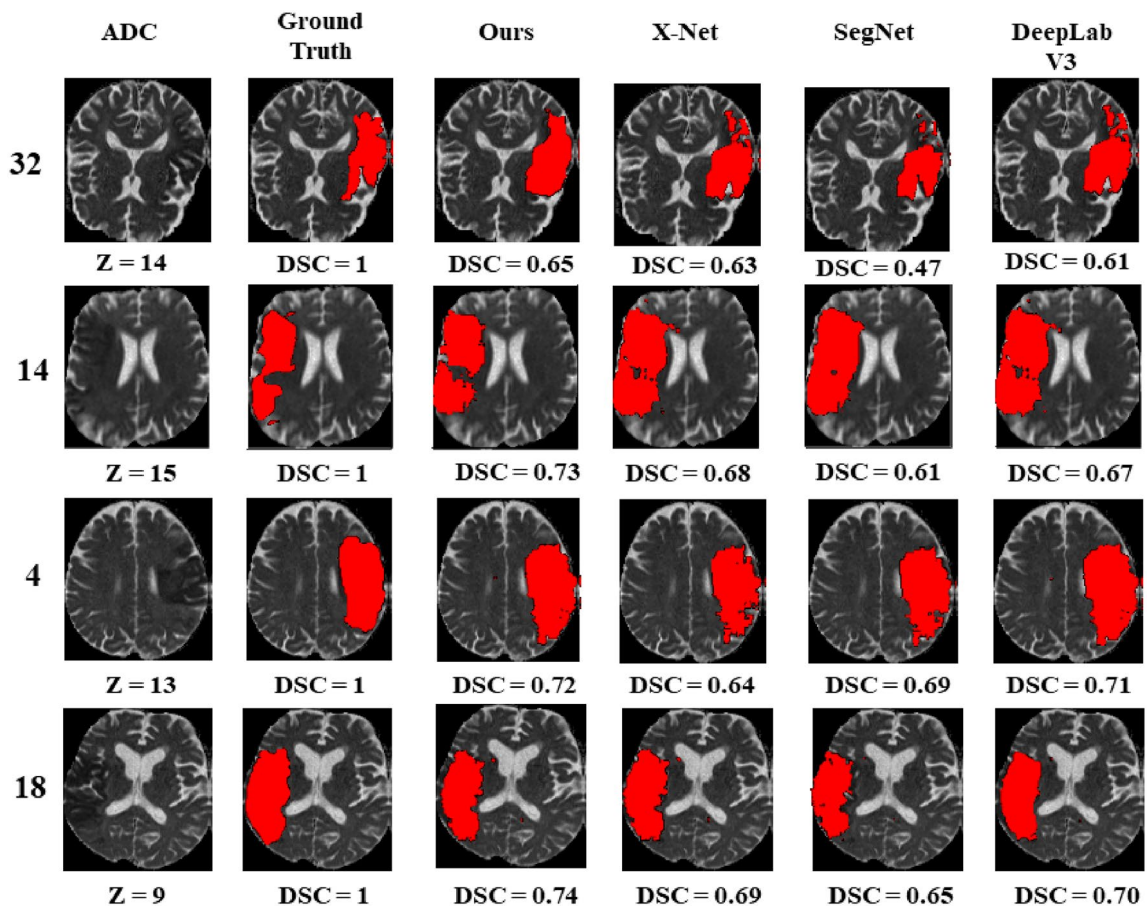

**Fig. 6** Visual Comparison of the segmentation results from our proposed method along with different existing methods. The first column represents the original input image of the Apparent Diffusion Coefficient (ADC) MRI sequence in 2D format along with their cut coordinate along x-xis included, the second column is the ground truth, the third column represents the segmentation map from our result. Along with visual comparison, we have also included the DSC scores to identify the accuracy of segmentation

**Table 3** Comparison of the proposed method with different existing methods based on the total number of model parameters and total training time on the ATLAS dataset with resized image dimension of $192 \times 192$

| Methods | Number of parameters | Training time |
|---|---|---|
| 2D UNet [32] | 7.77M | 6 h 20 min 13 s |
| 3D UNet [17] | 22.5M | 17 h 05 min 41 s |
| DeepLab V3 [8] | 21.3M | 16 h 44 min 15 s |
| Xnet [30] | 15.1M | 11 h 11 min 09 s |
| ResUNet [37] | 12.5M | 10 h 04 min 44 s |
| SegNet [1] | 14.7M | 10 h 55 min 37 s |
| **DFENet (proposed)** | **9.64M** | **8 h 23 min 12 s** |

We have also compared the total number of model parameters of our proposed method with different other existing frameworks in literature, shown in Table 3. It is evident from the table that our proposed method contains only 9.64M parameters as compared to 22.M parameters of 3D UNet. Existing state-of-the-art methods like SegNet, XNet, ResU-Net, and DeepLab V3 suffer from extremely high computational cost as compared to our proposed method with 14.7M, 15.1M, 12.5M, and 21.3M parameters respectively. UNet, with only 7.77M parameters, is lightweight as compared to DFENet, but produces poor segmentation map as shown in Table 2.

## Conclusion and Future Works

To gracefully address the shortcomings of existing segmentation methods, in this paper we present an end-to-end segmentation framework. The proposed DFENet can effectively extract information-rich contextual features and can predict accurate segmentation map by fusing 2D and 3D features. The proposed method is robust, effective and when compared to traditional segmentation models, outperforms them quite significantly, adding to its reliability and clinical importance. Therefore our main focus lies in developing a novel and effective segmentation framework. As it is clear from the discussion that our approach is suitable for volumetric image segmentation (as we are capturing 3D features also), and brain MRI being one such volumetric data, we have evaluated the model performance on this particular dataset. However, in future we would like to extend the experimentations on other datasets to analyze the model performance. In future, we plan to extend the model for semi-supervised segmentation to address the problem of the insufficient labelled dataset. This paper can be used as a testbed for further experimentations and the development of diverse segmentation frameworks for other biomedical applications as well.

**Funding** The authors did not receive support from any organization for the submitted work. No funding was received to assist with the preparation of this manuscript.

## Declarations

**Conflict of Interest** Hritam Basak declares that he has no conflict of interest. Rukhshanda Hussain declares that she has no conflict of interest. Ajay Rana declares that he has no conflict of interest.

**Ethical Approval** This article does not contain any studies with human participants or animals performed by any of the authors.